\documentclass{PoS}

\usepackage{epsfig}

\title{Running of the coupling and quark mass in SU(2) with two adjoint fermions}

\ShortTitle{SU(2) with two adjoint fermions}

\author{\speaker{Francis Bursa}\\
        University of Cambridge, UK\\
        E-mail: \email{fwb22@cam.ac.uk}}

\author{Luigi Del Debbio\\
        University of Edinburgh, UK\\
        E-mail: \email{luigi.del.debbio@ed.ac.uk}}

\author{Liam Keegan\\
        University of Edinburgh, UK\\
        E-mail: \email{liam.keegan@ed.ac.uk}}

\author{Claudio Pica\\
        University of Edinburgh, UK\\
        E-mail: \email{claudio.pica@ed.ac.uk}}

\author{Thomas Pickup\\
        University of Oxford, UK\\
        E-mail: \email{pickup@thphys.ox.ac.uk}}

\abstract{We simulate $SU(2)$ gauge theory with two massless Dirac
  fermions in the adjoint representation. We calculate the running of
  the Schr\"{o}dinger Functional coupling and the renormalised quark
  mass over a wide range of length scales. The running of the coupling
  is consistent with the existence of an infrared fixed point (IRFP),
  and we find $0.07 < \gamma < 0.56$ at the IRFP, depending on the
  value of the critical coupling.}

\FullConference{The XXVII International Symposium on Lattice Field Theory\\
		 July 26-31, 2009\\
		 Peking University, Beijing, China}

\begin{document}

\section{Introduction}
The symmetry breaking in technicolor theories
is communicated to the Standard Model by a further interaction at
some higher energy scale $M$. There is a tension on the
value of $M$: on the one hand $M$ needs to be large so that FCNC interactions are
suppressed, on the other it needs to be small to generate the heavier quark
masses. The effective operator for the quark masses is:
\begin{equation}
  \label{eq:qmass}
  \mathcal L = \frac{1}{M^2} \langle \Phi \rangle \bar\psi \psi\, ,
\end{equation}
where $\psi$ indicates the quark field, and $\Phi$ the field
which is responsible for electroweak symmetry breaking. In traditional technicolor
models $\Phi=\bar\Psi \Psi$ is the chiral
condensate of techniquarks. The
coefficient in Eq.~(\ref{eq:qmass}) is the chiral condensate at the scale
$M$:
\begin{equation}
  \label{eq:chrun}
  \left. \langle \bar\Psi\Psi \rangle \right|_M = 
  \left. \langle \bar\Psi\Psi \rangle \right|_\Lambda 
  \exp\left[\int_\Lambda^M \frac{d\mu}{\mu} \gamma(\mu)\right]\, ,
\end{equation}
where $\Lambda$ is the technicolor scale.
This suggests a possible way to resolve the tension
on $M$: if 
$\gamma$ is approximately constant and large over a sufficiently
long range in energies, then the condensate will be enhanced.
This scenario is known as {\it walking technicolor}. Gauge
theories with a large number of fermions, or with
fermions in higher--dimensional representations of
$SU(N)$~\cite{Sannino}, are candidates.
These theories could have a genuine IR
fixed point (IRFP), or simply be close to one.

The existence of an IRFP
is a difficult problem to address since it requires
quantitative computations in a strongly--interacting theory. Lattice
simulations provide first--principle results that can help in
determining the phenomenological viability of these models.
A number of
theories have been studied recently:
%~\cite{latt}: [Too many refs!]
SU(3) with 8, 10, 12 flavors of fermions
in the fundamental representation, SU(3) with fermions in the sextet
representation, and SU(2) with fermions in the adjoint
representation.
Existing simulations of the Schr\"odinger functional have identified a
possible fixed point in all the above--mentioned theories by noticing
a flat behaviour of the running coupling.

In this work we focus on SU(2) with 2 adjoint flavours,
and compute the running coupling in the SF scheme. We also compute
the running of the mass, and extract
the anomalous dimension.

\section{Basic formulation}
We define the running coupling $\overline{g}^2$ non-perturbatively
using the Schr\"{o}dinger Functional
method~\cite{Luscher:1991wu,Luscher:1992an}. This is defined on a
hypercubic lattice of size $L$, with boundary conditions chosen to
impose a background electric field on the system. The
spatial link matrices at $t=0$ and $t=L$ are set to:
\begin{equation}
  \label{eq:linkBC1}
  \left.U(x,k)\right|_{t=0}=\exp\left[\eta \tau_3 a/iL\right]\quad , \quad
  \left.U(x,k)\right|_{t=L}=\exp\left[(\pi-\eta) \tau_3a/iL\right]\, ,
\end{equation}
with $\eta=\pi/4$~\cite{Luscher:1992zx}. The fermion fields obey 
\begin{equation}
  \label{eq:ferm0}
  P_+\psi=0,~\overline{\psi}P_-=0~\mathrm{at}~t=0\quad , \quad
  P_-\psi=0,~\overline{\psi}P_+=0~\mathrm{at}~t=L\, ,
\end{equation}
where the projectors are defined as $P_\pm=1/2(1\pm\gamma_0)$. The
fermion fields also satisfy periodic spatial boundary
conditions~\cite{Sint:1995ch}.
We use the Wilson plaquette gauge action, and Wilson fermions in the
adjoint representation, as implemented in
Ref.~\cite{DelDebbio:2008zf}.

The coupling constant is defined as
\begin{equation}
  \label{eq:SFcoupling}
  \overline{g}^2=k \left< \frac{\partial S}{\partial \eta} \right>^{-1}
\end{equation}
with $k=-24L^2/a^2 \mathrm{sin}(a^2/L^2 (\pi-2\eta))$ chosen such that
$\overline{g}^2=g_0^2$ to leading order in perturbation theory. This
is a non--perturbative definition of the coupling which depends on
only one scale, $L$.

To measure the running of the quark mass, we calculate the
pseudoscalar density renormalisation constant $Z_P$. Following
Ref.~\cite{Capitani:1998mq}, $Z_P$ is defined by:
\begin{equation}
  \label{eq:ZPdef}
  Z_P(L)=\sqrt{3 f_1}/f_P(L/2)\, ,
\end{equation}
where $f_1$ and $f_P$ are the correlation functions involving the
boundary fermion fields $\zeta$ and $\overline{\zeta}$:
\begin{eqnarray}
  \label{eq:f1def}
  f_1&=&-1/12L^6 \int d^3u\, d^3v\, d^3y\, d^3z\,
  \langle
  \overline{\zeta}^\prime(u)\gamma_5\tau^a{\zeta}^\prime(v)
  \overline{\zeta}(y)\gamma_5\tau^a\zeta(z) 
  \rangle\, , \\
  \label{eq:fPdef}
  f_P(x_0)&=&-1/12 \int d^3y\, d^3z\,  \langle
  \overline{\psi}(x_0)\gamma_5\tau^a\psi(x_0)\overline{\zeta}(y)\gamma_5\tau^a\zeta(z)
  \rangle\, .
\end{eqnarray}
These correlators are calculated with the
spatial link matrices at $t=0$ and $L$ set to unity.

We run directly at $\kappa_c$, determined through the PCAC mass $m_{PCAC}(L/2)$, where
\begin{equation}
m_{PCAC}(x_0)=\frac{\frac{1}{2}(\partial_0+\partial_0^*)f_A(x_0)}{2f_P(x_0)}
\end{equation}
and
\begin{equation}
  f_A(x_0)=-1/12 \int d^3yd^3z \langle
  \overline{\psi}(x_0)\gamma_0\gamma_5\tau^a\psi(x_0)\overline{\zeta}(y)\gamma_5\tau^a\zeta(z) \rangle.
\end{equation}
Here $\partial_0$ and $\partial_0^*$ are
defined by $\partial_0f(x)=f(x+1)-f(x)$ and
$\partial_0^*f(x)=f(x)-f(x-1)$. The correlators are calculated on
lattices of size $L$ with the spatial link matrices at $t=0$ and
$L$ set to unity.

We define $\kappa_c$ by the point where $m_{PCAC}$ vanishes on the $6^4$ and $8^4$
lattices, a linear
extrapolation in $a/L$ from these values, and the values for $16^4$
lattices quoted in Ref.~\cite{Hietanen:2009az}.
In practice we achieve $|a m_\mathrm{PCAC}|\lesssim 0.005$.
We check explicitly that there is no residual sensitivity to the small
remaining quark mass by repeating some of our simulations at
$m_\mathrm{PCAC} \sim 0.02$.

\section{Evidence for fixed points}
Recent lattice studies have focused on the running of the SF coupling, emphasizing the slow
running of this
quantity~\cite{Appelquist:2007hu,Shamir:2008pb,Appelquist:2009ty,Hietanen:2009az}. These results have to be interpreted with
care. Lattice data can find a range of energies over which no running is
observed, but one cannot conclude that this extends to arbitrarily large distances, as one would expect in the
presence of an IRFP. On the other hand, if the plateau has a finite extent, {\it i.e.} if
the theory seems to walk, the behaviour of the running coupling
depends on the scheme, and therefore the conclusions
become less compelling.

There are instances where the beta function of an asymptotically free
theory is numerically small. This is the case of the theory considered
in this work in the perturbative regime. In this case, even
though the theory does not have a fixed point, the running of the coupling is very
slow. 
High accuracy is needed in order to resolve a ``slow''
running; therefore numerical studies of potential IRFP need high statistics, and 
a robust control of systematics. In particular it is important to extrapolate to the continuum limit to eliminate
lattice artefacts.

Studies of the SF running coupling are a useful tool to expose the possible
existence of theories that show a conformal behaviour at large distances. However the
results have to be interpreted with care; they are unlikely to
provide conclusive evidence about the existence of a fixed point by
themselves. A more convincing picture can emerge
when they are combined with spectral studies.

\section{Results for the coupling}
\begin{figure}[h]
  \centering
  \epsfig{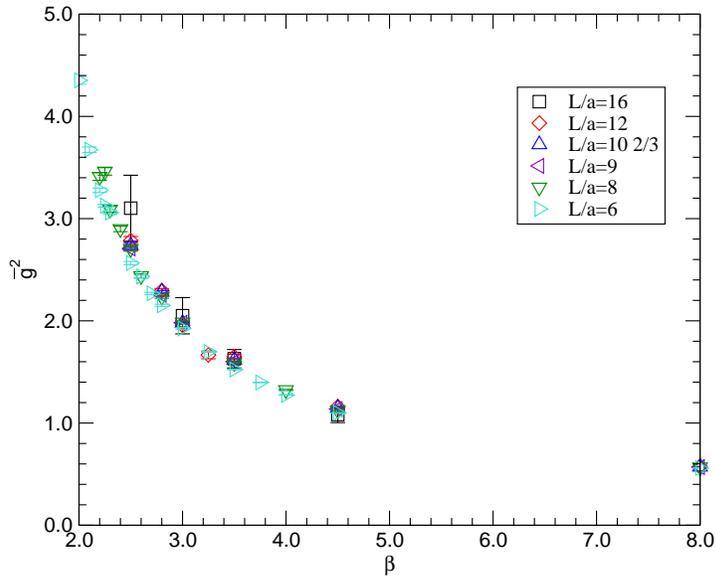}
  \caption{Data for the running coupling as computed from lattice
    simulations of the Schr\"odinger functional.}
  \label{fig:SFdata}
\end{figure}

We have measured the coupling $\overline{g}^2(\beta,L)$ for a range of
$\beta,L$. Our results are plotted in Fig.~\ref{fig:SFdata}. 
They are directly comparable to those of
Ref~\cite{Hietanen:2009az}, and agree within statistical errors.
It is clear that the coupling is very similar for different $L/a$ at a given
value of $\beta$, and hence that it runs slowly.

In order to see how the coupling runs we define the step scaling
function $\sigma(u,s)$ as
\begin{equation}
\label{eq:sigma}
\sigma(u,s) = \overline{g}^2(sL)|_{\overline{g}^2(L)=u}
\end{equation}
This is the continuum extrapolation of $\Sigma(u,s,a/L)$ which we
calculate at various $a/L$.

We interpolate quadratically in $a/L$ to find values of
$\overline{g}^2(\beta,L)$ at $L=9,10\frac{2}{3}$, which gives us data
for four steps of size $s=4/3$ for $L\rightarrow sL$: $L=6,8,9,12$;
$sL=8,10\frac{2}{3},12,16$. Then for each L we perform an
interpolation in $\beta$.
We can then find estimates of $\Sigma(u,4/3,a/L)$ at any $u$. A
continuum extrapolation is then performed in $a/L$
to give an estimate of $\sigma(u)\equiv\sigma(u,4/3)$. The $L=6$
data was found to have large $O(a)$ artifacts, and we have too few
$L=16$ points to constrain the interpolation functions, so neither are
used in the continuum extrapolation.

\begin{figure}[h]
  \centering
  \epsfig{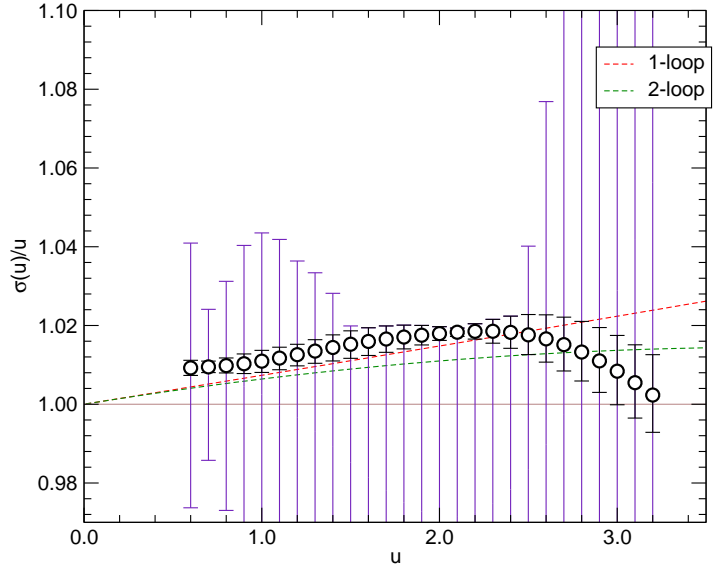}  
  \caption{The relative step--scaling function $\sigma(u)/u$. The
    black circles use only a constant continuum extrapolation. The
    purple error bars
    include both constant and linear continuum extrapolations.  }
  \label{fig:SFlinsigma}
\end{figure}

The resulting values for $\sigma(u)$ can be seen in
Fig.~\ref{fig:SFlinsigma}. The systematic errors
from varying the interpolation
functions or the continuum extrapolation were significantly
larger than the statistical errors.
To quantify this, we recalculated $\sigma(u)$ with a range of different interpolation and
extrapolation functions. The resulting extremal values of $\sigma(u)$
were used as upper and lower bounds on the central value. The black
error bars
in Fig.~\ref{fig:SFlinsigma} are determined in this way, but using
only a constant continuum extrapolation. These values are consistent
with a fixed point in the region $\overline{g}^2 \sim 2.0-3.2$, as reported
in Ref.~\cite{Hietanen:2009az}. 
The errors from also including the linear continuum extrapolation are
much larger and mask any evidence for a fixed point, as also shown in
Fig.~\ref{fig:SFlinsigma}.

%\begin{figure}[h]
%  \centering
%  \epsfig{file=section4/sigma/const_sigma.eps,scale=0.45,clip}  
%  \caption{The relative step--scaling function $\sigma(u)/u$ obtained
%    after extrapolating the lattice data to the continuum limit using
%    a constant extrapolation (i.e. ignores lattice artifacts). Note
%    that $\sigma(u)/u=1$ identifies a fixed point.  }
%  \label{fig:SFconstsigma}
%\end{figure}

\section{Running mass}
\begin{figure}[h]
  \centering
  \epsfig{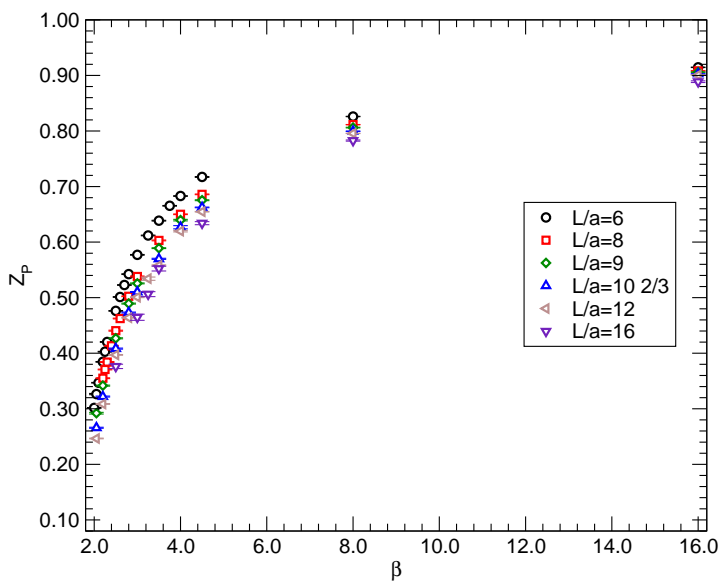}  
  \caption{Data for the renormalisation constant $Z_P$
    from lattice simulations of the Schr\"odinger
    functional.}
  \label{fig:ZPdata}
\end{figure}

We have measured
$Z_P(\beta,L)$ for a range of $\beta,L$.  We plot our results in
Fig.~\ref{fig:ZPdata}, where we see a clear trend in
$Z_P$ as a function of $L$ at all $\beta$.
The step scaling function $\sigma_P(u,s)$ is then defined
as:
\begin{equation}
\label{eq:sigma_p}
\sigma_P(u,s) = \left
  . {\frac{Z_P(sL)}{Z_P(L)}}\right | _{\overline{g}^2(L)=u}
\end{equation}
We extract this from a continuum extrapolation of
$\Sigma_P(u,s,a/L)$.

The method for calculating $\sigma_P(u)\equiv\sigma_P(u,4/3)$ is
similar to that for calculating
$\sigma(u)$. $Z_P$ converges faster than
$\overline{g}^2$ and we have better $16^4$ data so here we use 4 points
in our continuum extrapolations. Again the errors are dominated by
systematics, in particular the
choice of continuum extrapolation. We find good agreement
with the 1-loop perturbative prediction.
%as shown in Fig.~\ref{fig:sigmaP}.

%\begin{figure}[h]
%  \centering
%  \epsfig{file=section5/sigma_P/sigma_P.eps,scale=0.45,clip}  
%  \caption{The step-scaling function for the running mass
%    $\sigma_P(u)$, using a linear continuum extrapolation. The grey
%    error bars come from also including a constant extrapolation of
%    the two points closest to the continuum, and give an idea of the
%    systematic error in the continuum extrapolation.}
%  \label{fig:sigmaP}
%\end{figure}

\begin{figure}[h]
  \centering
  \epsfig{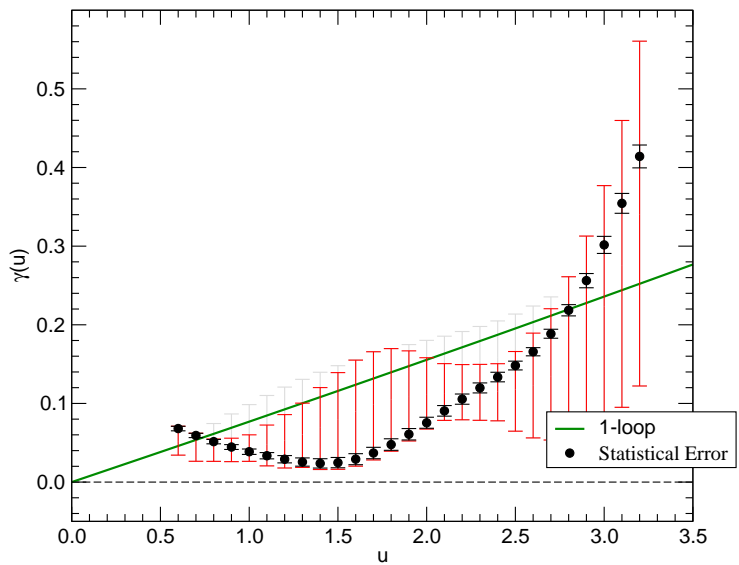}  
  \caption{The mass anomalous dimension $\gamma(u)$. Red
error bars use only a linear continuum extrapolation. Grey error
    bars include both constant and linear continuum extrapolations.}
  \label{fig:gamma}
\end{figure}

We cannot determine directly the running of the mass with scale
since we observe no running of the coupling within errors.
However, we can define an estimator for the anomalous dimension,
\begin{equation}
\label{eq:tau}
\gamma(u) = -\frac{\ln\left|\sigma_P(u,s)\right|}{\ln\left|s\right|},
\end{equation}
which is equal to the anomalous dimension at an IRFP, and which
we plot in Fig.~\ref{fig:gamma}. We see that
it is rather small over the range of interest; in particular,
at $\overline{g}^2=2.2$, the benchmark value for the IRFP in~\cite{Hietanen:2009az},
we have $\gamma=0.114^{+78}_{-35}$, and over the whole range
$\overline{g}^2=2.0-3.2$ consistent with an IRFP
in~\cite{Hietanen:2009az}, we find $0.07 < \gamma < 0.56$.

\section{Conclusions}
In these proceedings we have presented results for the running of the
Schr\"{o}dinger Functional coupling $\overline{g}^2$ and the mass
anomalous dimension $\gamma$.

Our results for the running of the coupling are
completely consistent with those of Ref.~\cite{Hietanen:2009az}. Our
statistical errors are larger; however, we have carried out our
analysis in a way that allows us to take the continnum limit with full
control over the resulting systematic errors. Our results appear to
show a slowing in the running of the coupling above $\overline{g}^2=2$
or so, and are consistent with the presence of a fixed point
at somewhat higher $\overline{g}^2$. However, once we
include the systematic errors from the continuum extrapolation
that our results no longer give any evidence for a fixed point. 

By contrast, we find that the behaviour of the anomalous dimension
$\gamma$ is much easier to establish. We find
a moderate anomalous dimension, close to the 1-loop perturbative
prediction, throughout the range of $\beta$ explored. 
In particular,
at $\overline{g}^2=2.2$, the benchmark value for the IRFP in~\cite{Hietanen:2009az},
we find $\gamma=0.114^{+78}_{-35}$. This value is much smaller than that
required for phenomenology, which is typically of order 1-2. Such large values
of $\gamma$ are clearly inconsistent with our results.

The anomalous dimension is more vital than the running of
$\overline{g}^2$ for phenomenology; if it is not large then the
presence or absence of walking behaviour becomes academic. Hence the
implications of our measurement of $\gamma$ for
minimal walking technicolor deserve to be studied carefully.

The results presented here are preliminary, and the systematic errors
need to be reduced to make our conclusions more robust.
Using larger lattices would 
make the continuum extrapolations more accurate, and it may also be
necessary to use an improved action to reach the
precision required to show the existence of an IRFP or of walking behaviour.
However this is very unlikely to affect our
phenomenologically most important result, namely that $\gamma$ is not large.

\end{document}